%% file: riken.tex
\documentstyle[12pt,epsfig]{article}
\input{defi}
\begin{document}
\begin{titlepage}
\begin{flushright}
TAUP\,\,2450 - 97\\
DESY \,\,97 - 171\\
September 1997\\
\end{flushright}

\vspace{4cm}
\begin{center}
\begin{Large}
{\bf WHERE ARE THE BFKL POMERON }\\
{\bf and}\\ 
{\bf SHADOWING CORRECTIONS IN DIS ?           }
 
\end{Large}
\end{center}
\vspace{1cm}
 \centerline{\large Eugene  Levin}
\centerline{\it School of Physics and Astronomy, Tel Aviv University}
\centerline{\it  Ramat Aviv, 69978, ISRAEL}
\centerline{\it and}
\centerline{\it DESY Theory, Notkestr. 85, D - 22603, Hamburg, GERMANY}
~
\centerline{{\tt leving@ccsg.tau.ac.il; \,\,levin@mail.desy.de;}}
\centerline{}
\centerline{}

{\it Talk given at the RIKEN BNL WS on `` Perturbative QCD as a Probe
of Hadron Structure", BNL LI July 14 - 25,1997.}
\centerline{}
\centerline{}
\centerline{}
\centerline{}

{\bf Abstract:} 
In this talk,
I will argue that the HERA experimental data show that the typical
parameter ($\kappa$) responsible for the value of the shadowing
corrections (SC) in DIS is so large that the BFKL Pomeron is hidden under
SC.
The SC turn out to be large enough but mostly for the gluon structure
function which is not well determined by the available experimental data
and by the current  theoretical procedure. 

\end{titlepage}

In this talk I am going to answer two questions:

{\bf Q1:} Where is the BFKL \cite{BFKL} Pomeron?

{\bf Q2:} Where are shadowing corrections (SC)?

Actually, the answers have been presented in our paper  \cite{AGLP}, but
here I will discuss them in more details.
\centerline{}
First, let me explain why it is reasonable to ask such questions. Indeed,
at first sight, the situation looks very transparent, namely, the HERA
data can be described by means of the usual DGLAP \cite{DGLAP} evolution
equations without any other ingredients such as the BFKL Pomeron and / or
SC ( see any of plenary talks during the past three years). 
My personal opinion is that this fact brought more questions
than answers since we need to show ( to justify theoretically our
approach)  that the corrections due to the BFKL
dynamics and/or due to the SC   are negligible small at least at the HERA
kinematic region. If it is not so ( as I will show below) the DGLAP
approach is not better or worse than any other model developed to
describe the
experimental data. The main goal of this talk is to show that the
experimental data from HERA confirm  that both the BFKL contribution and
the
SC should be rather large in the HERA kinematic region.

Actually, everything that I want to tell is given in Fig.1, but I need to
explain what are  plotted in this figure.
\centerline{}
\centerline{}
{\bf 1.} $< \gamma >\,=\,\frac{1}{2}$ and $< \gamma >\,=\,1$.
\centerline{}
 Let me recall a standard procedure
of solving of the DGLAP evolution equations.  

The first step: we introduce
moments of the structure function, namely,  
$$
x G(x,Q^2)\,=\,\frac{1}{2\pi i}\int_C e^{-\omega \,\ln(1/x)} \,M(\omega,
Q^2) \,d \omega,
$$
 where
contour
$C$ is located to the right of all singularities of moment $M(\omega,
Q^2)$.

 The second step: we find the solution to the DGLAP equation
for moment
\beq \label{1}
\frac{d M(\omega, Q^2)}{d \ln Q^2}\,\,=\,\,\gamma(\omega)\,M(\omega,
Q^2)\,\,.
\eeq
The solution is
\beq \label{2}
M(\omega, Q^2)\,\,=\,\,M(\omega,Q^2_0)\,
\cdot\,e^{\gamma(\omega)\,\ln(Q^2/Q^2_0)}\,\,.
\eeq
Here $M(\omega,Q^2_0)$ is the nonperturbative input which should be taken
from experimental data or from ``soft" phenomenology ( model).

The third step: we find the solution for the parton structure function
using the inverse transform, namely:
\beq \label{4}
xG(x,Q^2) \,\,=\,\,\int_C \,\frac{d \omega}{2 \pi i}\,\,e^{\omega
\,\ln(1/x)\,+\,\gamma(\omega)\,\ln(Q^2/Q^2_0)}\,M(\omega,Q^2_0)\,\,. 
\eeq
Therefore, to find a solution of the DGLAP equation we need to know the
nonperturbative input $M(\omega,Q^2_0)$ and the anomalous dimension
$\gamma(\omega)$, which we can calculate in perturbative QCD. The
anomalous dimension $\gamma(\omega)$ has been calculated in pQCD and the
result of calculations can be written in the form:
\beq \label{5}
\gamma(\omega)\,\,=\,\,\gamma^{BFKL}( \frac{\alpha_S N_c}{\pi\,\omega})\,\,-\,\,
\frac{\alpha_S N_c}{\pi\,\omega}\,\,+\,\,\alpha_S\,\gamma_1
(\omega)\,\,+\,\,\alpha^2_S\,\gamma_2 (\omega)\,\,,
\eeq
where both functions $\gamma_1$ and $\gamma_2$ are known as well as
$\gamma^{BFKL}$. 
  Using \eq{5} we can  discuss  what has been done in the global fits
\cite{GFIT}. The
value of the
anomalous dimension has been calculated in $\as$ and  $\alpha^2_S$ orders 
( two
last terms in Eq.(5)) and the nonperturbative input has been taken in
the form  $M(\omega,Q^2_0)\,\propto \,\frac{1}{\omega - \omega_0}$ with $
\omega_0\,\approx\,$ 0.2 - 0.3.   This means that the structure function
at $Q^2 = Q^2_0$ increases as $ x^{- \omega_0}$ at $x \rightarrow 0$
\footnote{Strictly speaking this statement is correct for two global
fits: MRS and CTEQ. The GRV fit  has a  different initial condition,
namely, the evolution has been started at very low value of $Q^2$ but
with the initial distribution which is flat at low $x$ .}. 
However, one can see that the $\gamma^{BFKL}$ should be essential in the
region of low $x$ where $\omega \,\rightarrow\,0$ since
\beq \label{6}
\gamma^{BFKL}(\frac{\alpha_s N_c}{\pi\,\omega})\,\,=\,\,\frac{\alpha_s
N_c}{\pi\,\omega}\,\,+\,\,\sum_{n = 4}\,C_n \,(\,\frac{\alpha_s
N_c}{\pi\,\omega}\,)^n\,\,\rightarrow \,|_{\omega\,\rightarrow\,0} 
\,\,\frac{1}{2} \,\,+\,\,\sqrt{\frac{\omega\,-\,\omega_L}{\Delta}}\,\,,
\eeq
where $\omega_L$ and $\Delta$ have been calculated \cite{BFKL}.

This equation reflects the main properties of the BFKL Pomeron: the
limited value of the anomalous dimension and the importance of all terms
of the order of $(\frac{\alpha_s N_c}{\pi\,\omega})^n$ in the region of
small $\omega$.  All attempts to estimate the values of the  BFKL
terms in the anomalous dimension  \cite{EKL} show that they are essential
in the HERA kinematic region. Here, we choose a different way of
presentation of this well known fact, namely, we introduce average
anomalous dimension 
 $<\gamma >$ which is equal to
\beq \label{7}
<\gamma >\,\,=\,\,\frac{1}{x G(Q^2,x)}\cdot\frac{\partial x
G(Q^2,x)}{\partial \ln(Q^2/Q^2_0)}\,\,=\,\,\frac{\int_C \,\frac{
\gamma(\omega)\,\,d
\omega}{2\pi i}\,\,e^{\omega\,\ln(1/x)\,+\,\gamma(\omega)\,
\ln(Q^2/Q^2_0)}\,M(\omega,Q^2_0)}{\,\int_C \,\frac{d \omega}{2 \pi
i}\,\,e^{\omega
\,\ln(1/x)\,+\,\gamma(\omega)\,\ln(Q^2/Q^2_0)}\,M(\omega,Q^2_0)}\,\,.
\eeq
Function $<\gamma >$ describes the behaviour of the anomalous dimension
quite well since at low $x$ the deep inelastic structure functions can be 
calculated in the semiclassical approach \cite{GLR}  in which, for example
$x G(Q^2,x)$, is equal to
\beq \label{8}
x G(Q^2,x)\,\,=\,\,C(\ln(1/x),\ln(Q^2/Q^2_0))\,\cdot\,\frac{1}{x^{<
\omega(\ln(1/x),\ln(Q^2/Q^2_0))>}}\,\cdot\,(
\frac{Q^2}{Q^2_0})^{<\gamma(\ln(1/x),\ln(Q^2/Q^2_0))>}\,\,
\eeq
where functions $C$,$<\omega>$ and $<\gamma>$ are smooth function of
$\ln(1/x)$ and $\ln(Q^2/Q^2_0)$.

In Fig.1 we plotted two lines with   $<\gamma>$\, =\, $\frac{1}{2}$ and
$<\gamma>$\,=\,1 . We expect a large the BFKL contribution in  the
kinematic region between these two lines  and one can see in Fig.1 that 
we have penetrated this region at HERA.
    
\begin{figure}[htbp]
\centerline{\epsfig{file=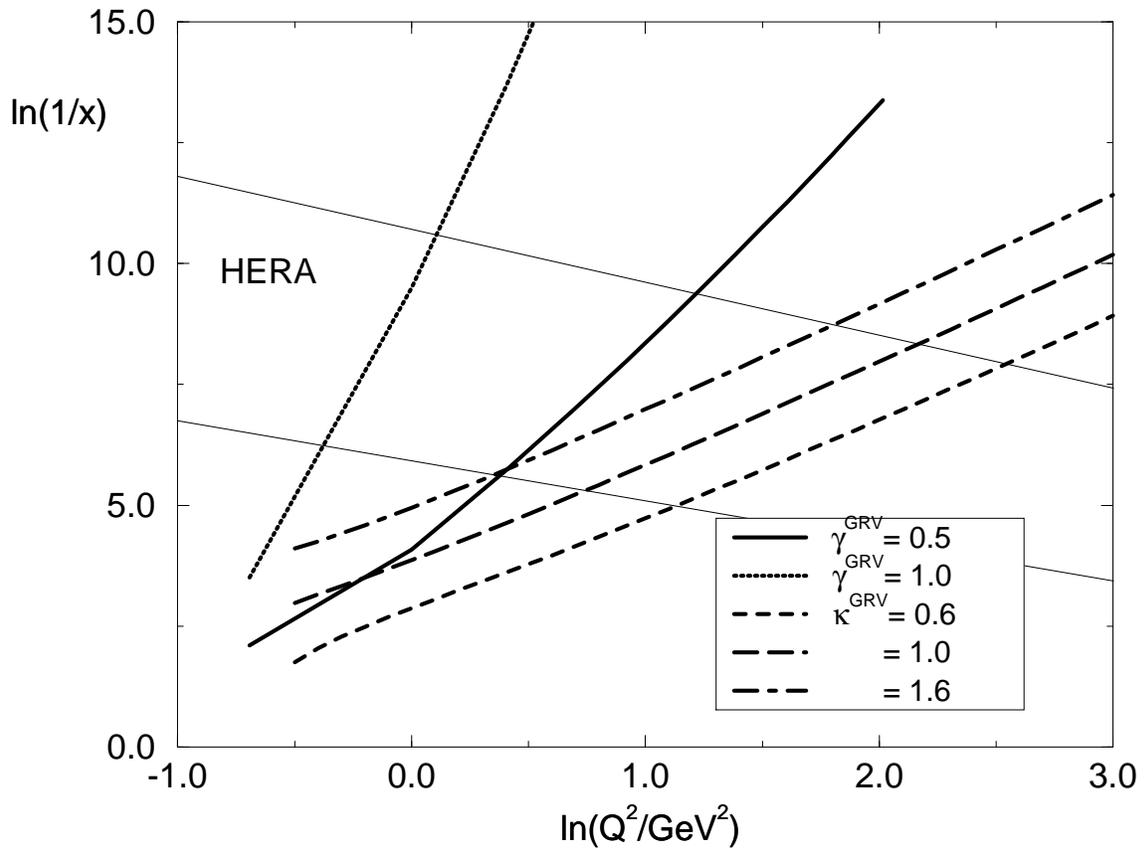,width=165mm}}
\caption{Contours for $ <\gamma > $ = 1 and  1/2 and $\kappa$ =
0.6,1,1.6   for
the GRV95 gluon density  and HERA kinematic region.}
\label{Fig.1}
\end{figure}   

\centerline{}
{\bf 2.} $\kappa$.
\centerline{}
 From  HERA data we can evaluate also
the probability $\kappa$ of the parton - parton (gluon - gluon)
interaction,
which is given by \cite{GLR}, \cite{MUQI}
\beq
\label{9}
\kappa\,\,=\,\,x G(x,Q^2) \frac{\sigma(GG)}{ Q^2 \, \pi R^2}\,\,=\,\,
\frac{ 3 \,\pi\,\alpha_S}{ Q^2\,R^2}\,\,xG(x,Q^2)\,\,,
\eeq
where $xG(x,Q^2)$ is the number of partons ( gluons) in the parton cascade
and $R^2$ is the radius
 of the area populated by gluons in a nucleon. $\sigma (GG)$ is the gluon   
cross section inside the parton cascade  and
was evaluated in \cite{MUQI}.

The observation is that we know from the HERA data both the value of the
gluon structure function and the value of $R^2$ in \eq{9}. Indeed, the
available parameterizations such as MRS, CTEQ and GRV \cite{GFIT} give the
value of the gluon structure function 
with sufficiently large differences
in its value. However, this difference is less than  50\% and becomes
smaller 
with improvement of the experimental data on $F_2(x,Q^2)$  ( see Fig.2 ).

\begin{figure}[htbp]
\begin{tabular}{c c}
\psfig{file=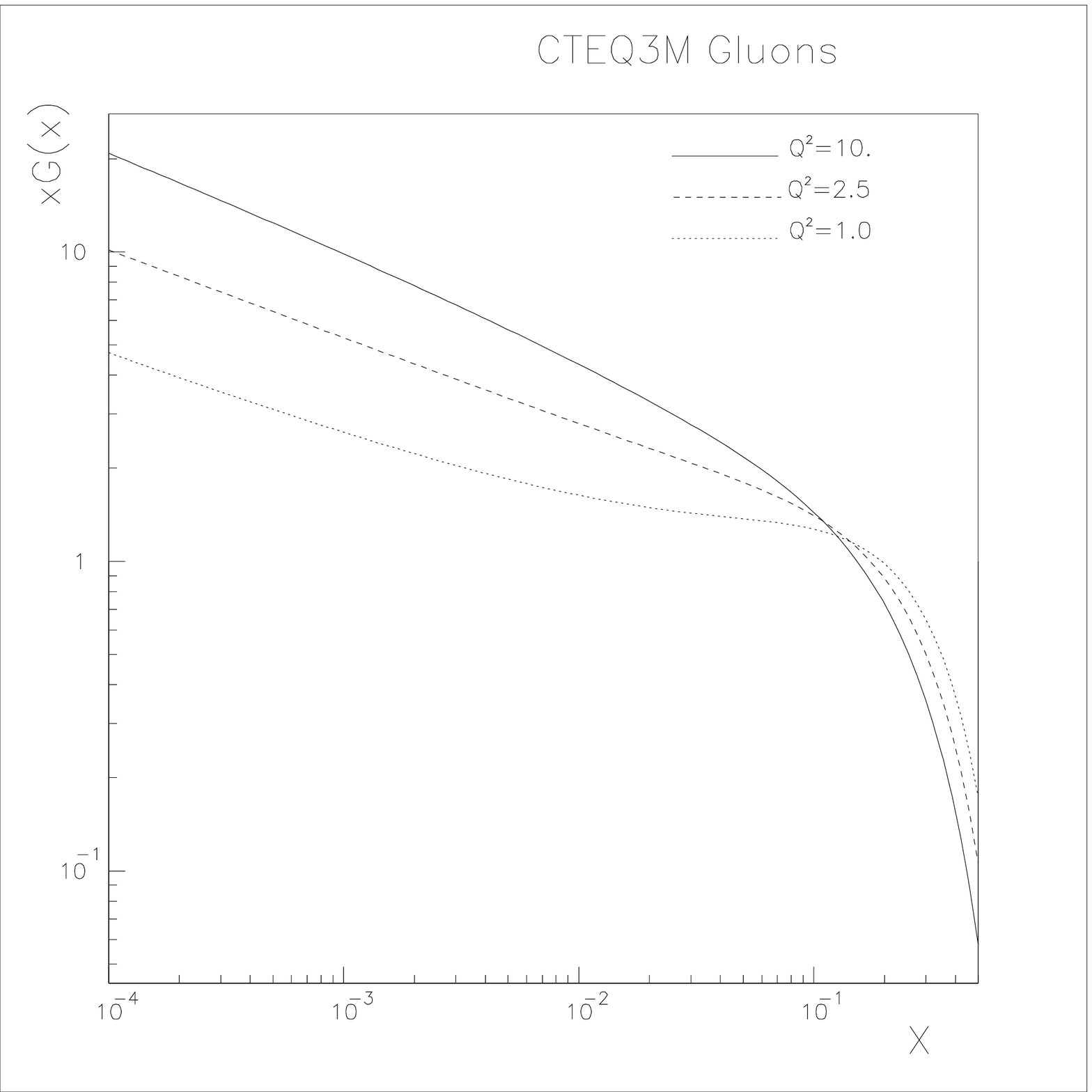,width=70mm} &
\psfig{file=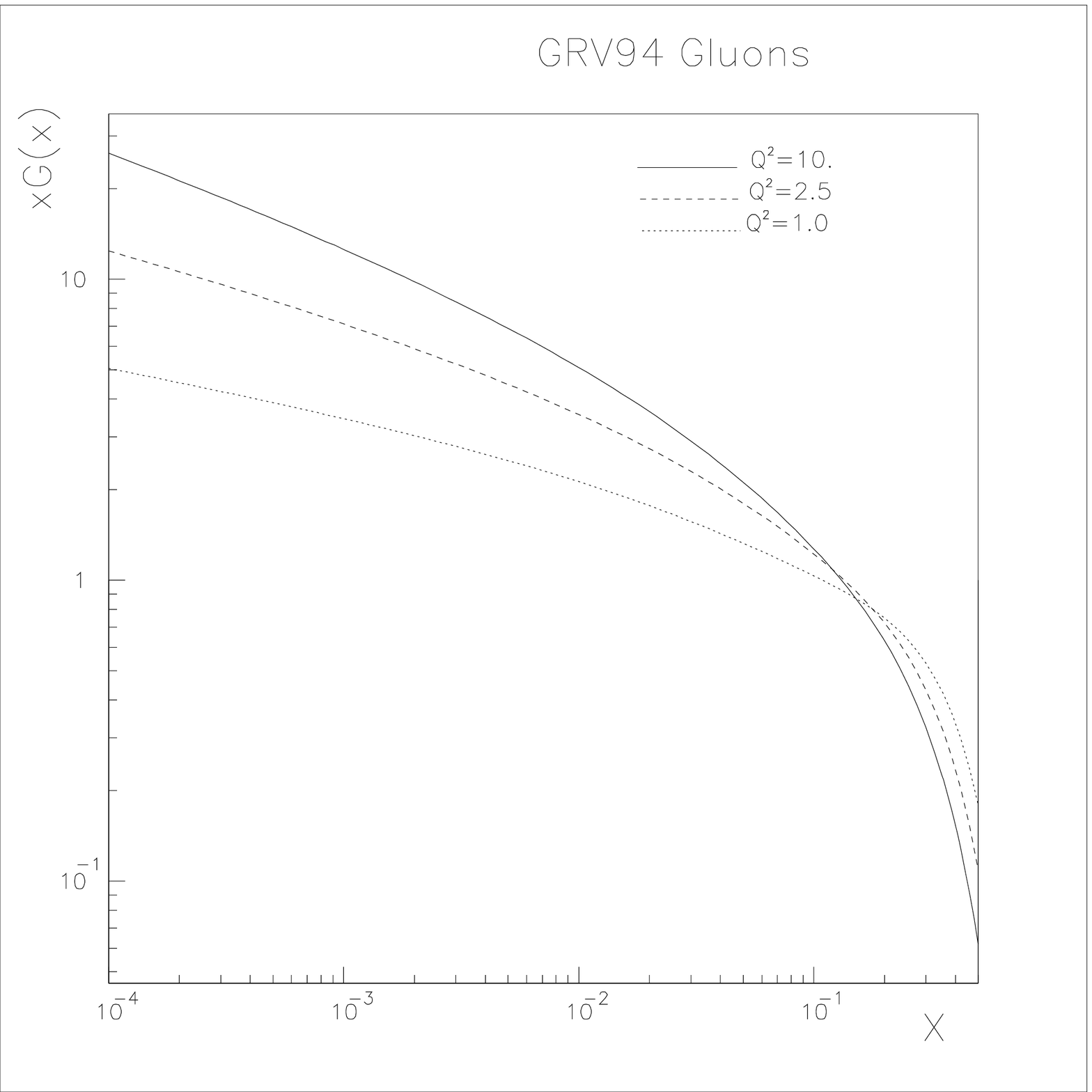,width=70mm}
\\
\psfig{file=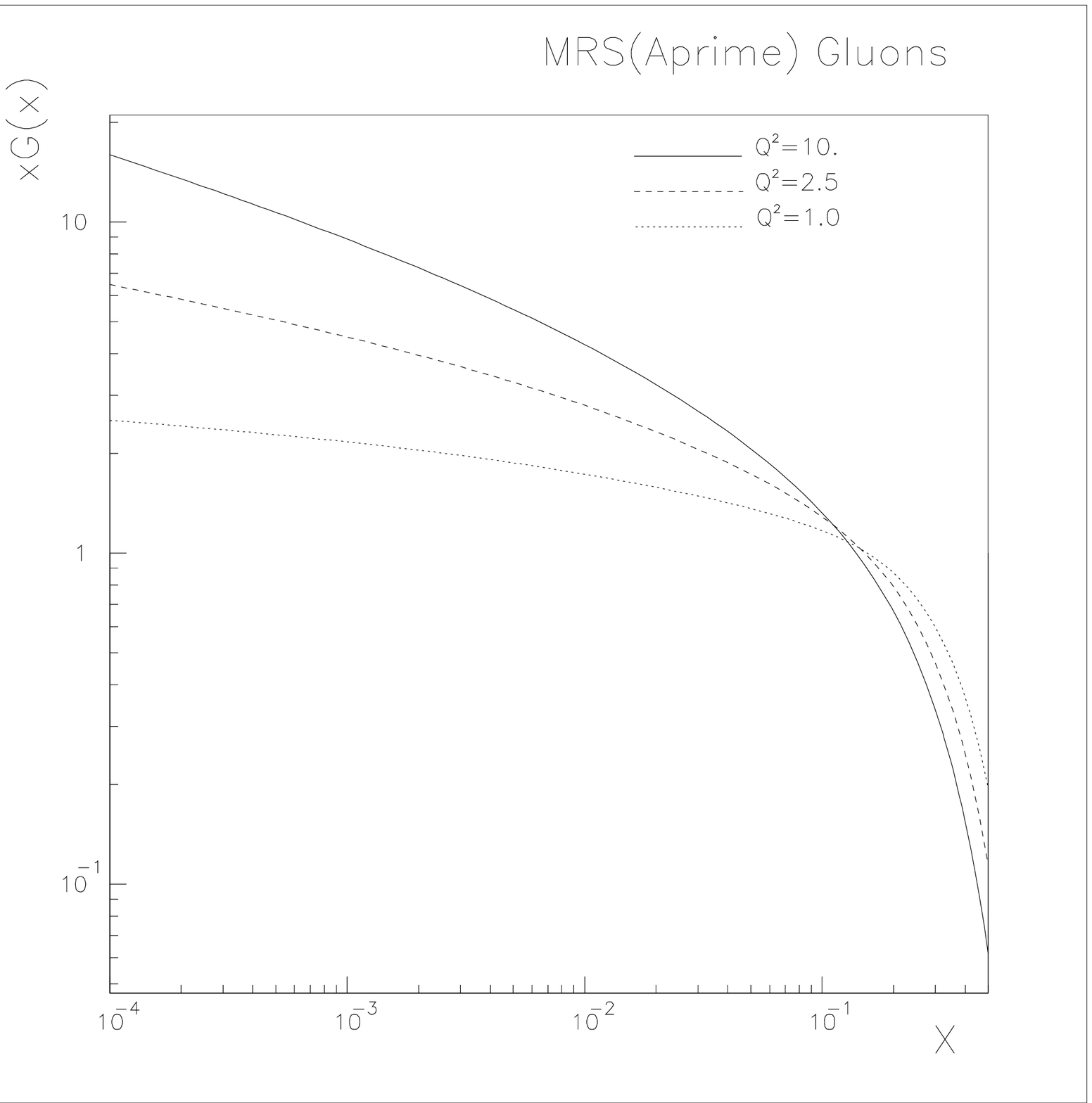,width=70mm} &
\psfig{file=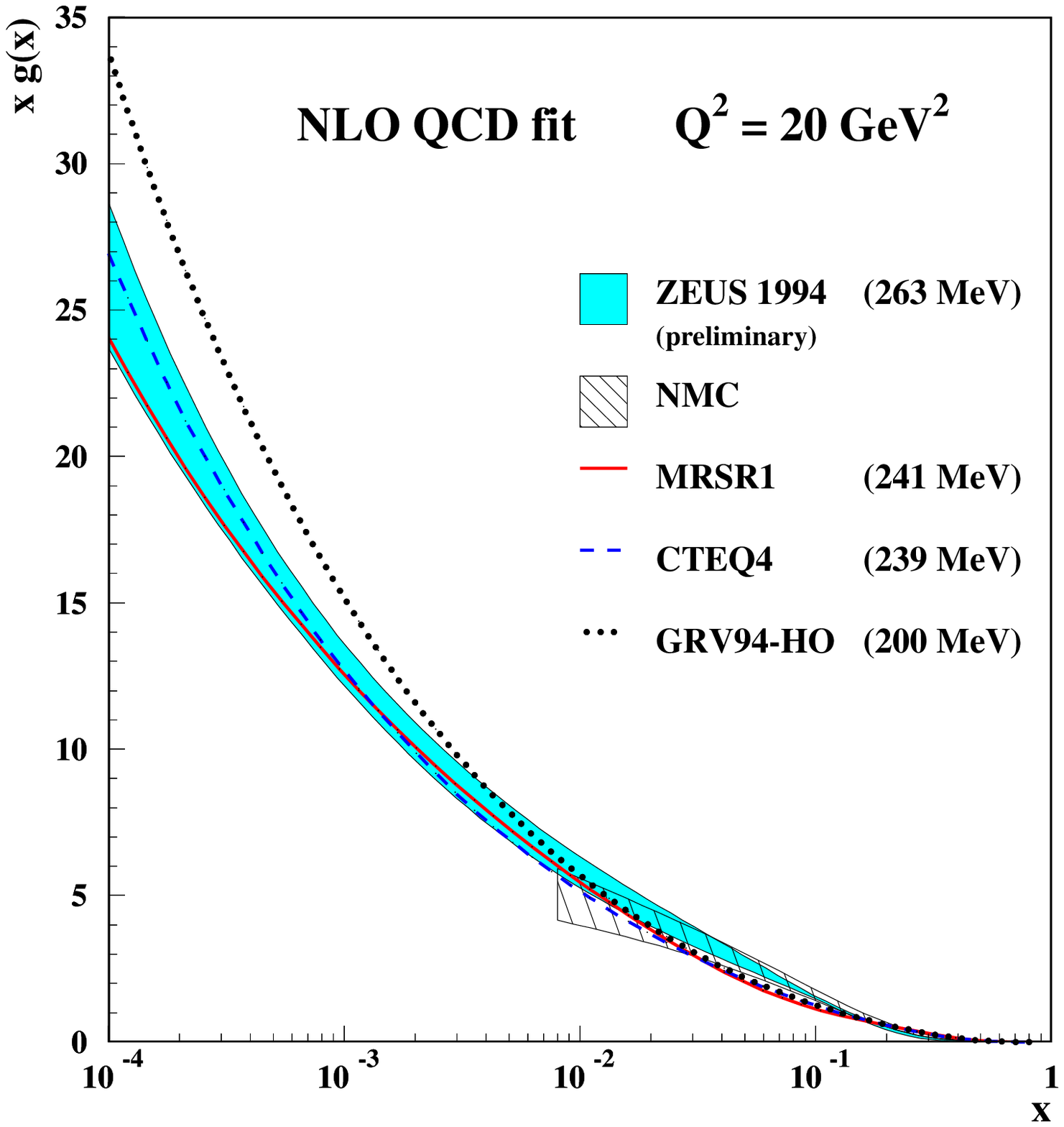,width=70mm,height=60mm}
\\
\end{tabular}
\caption{The value of the gluon structure function in different
parameterizations.}
\label{Fig.2}
\end{figure}

The most important and new information is the fact that   
using HERA
data on photoproduction of J/$\Psi$ meson
\cite{HERA}  the
value of $R^2$ can be  estimated as $R^2 \,\,\leq\,\,5 \, GeV^{-2}$
\cite{SLOPE}.  

\begin{figure}[hbtp]
\epsfig{file=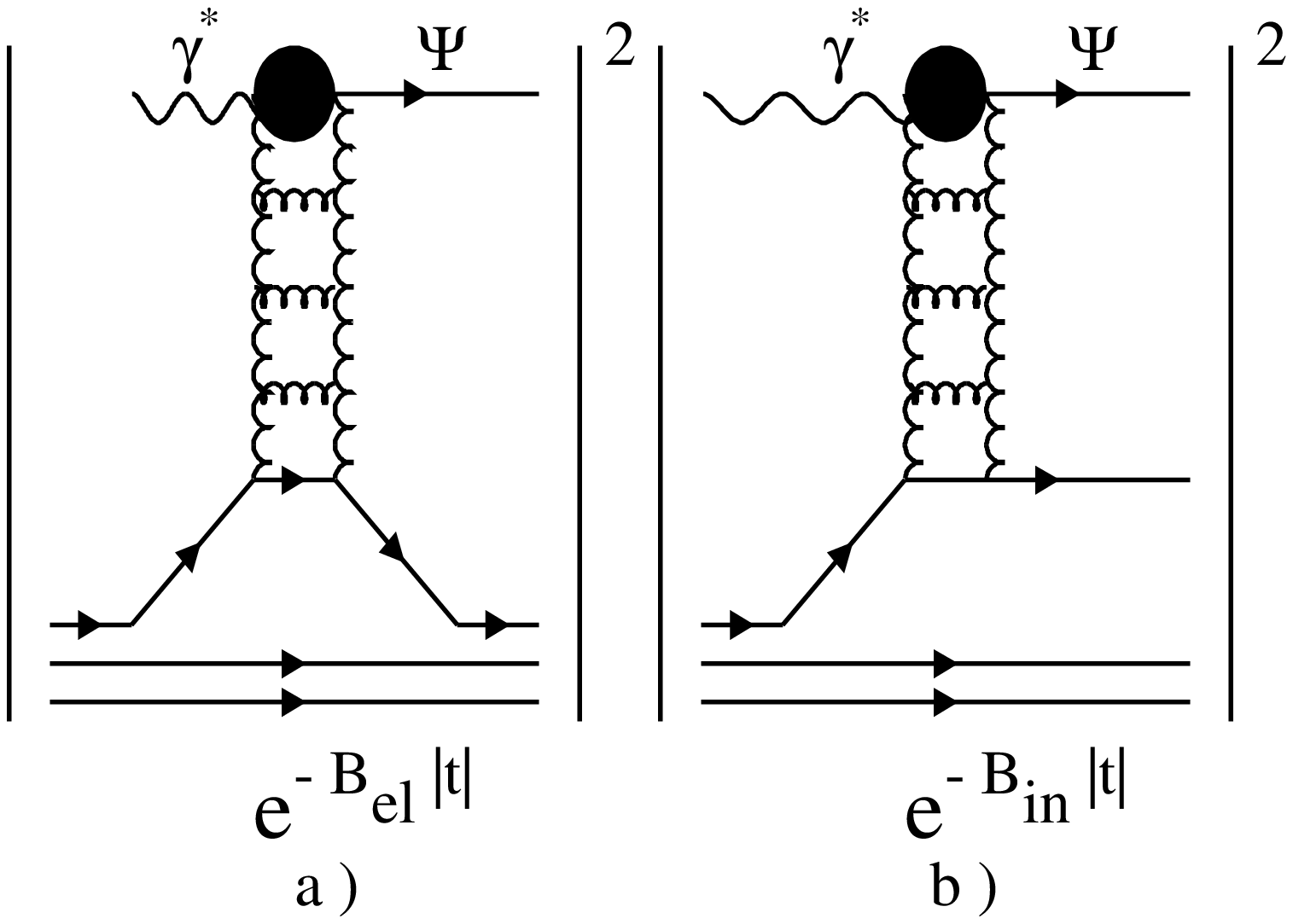,width=140mm}
\caption{The J/$\psi$ production without (a) and with ( b)
proton dissociation.}
\label{Fig.3}
\end{figure}

Indeed, (i) the experimental values for the slopes ( see
Fig.3 )
are $B_{el}\,=\,4 \,GeV^{-2}$ and $B_{in}\,=\,1.66\, GeV^{-2}$  and (ii)
the cross section for J/$\Psi$ production with and without proton
dissociation are equal \cite{HERA}.  Taking into account both facts we can
estimate the value of $R^2$ ( see Ref.\cite{SLOPE} for details) which
appears in calculation of the SC
(Glauber corrections) due to integration over the momentum transferred
($q^2_{\perp}\, =\, |t|$)
along the gluon ladders (see Fig.4) neglecting $t$ dependence of the upper
vertex in Fig.4.

In Fig.3 we show the picture for the diffractive production of J/$\Psi$ in
the additive quark model, in which two radii naturally appear as the
radius of the hadron and a proper radius of the constituent quark. In all
our estimates we did not need this particular model but it is interesting
to mention that our estimates give the same value of the average radius
as in the additive quark model.

It should be stressed that such an estimate gives the value for $R^2$
which lead to the value of the cross section for the double parton
scattering
 measured by the  CDF collaboration  at the Tevatron \cite{CDF}.
  
\begin{figure}[htbp]
\epsfig{file=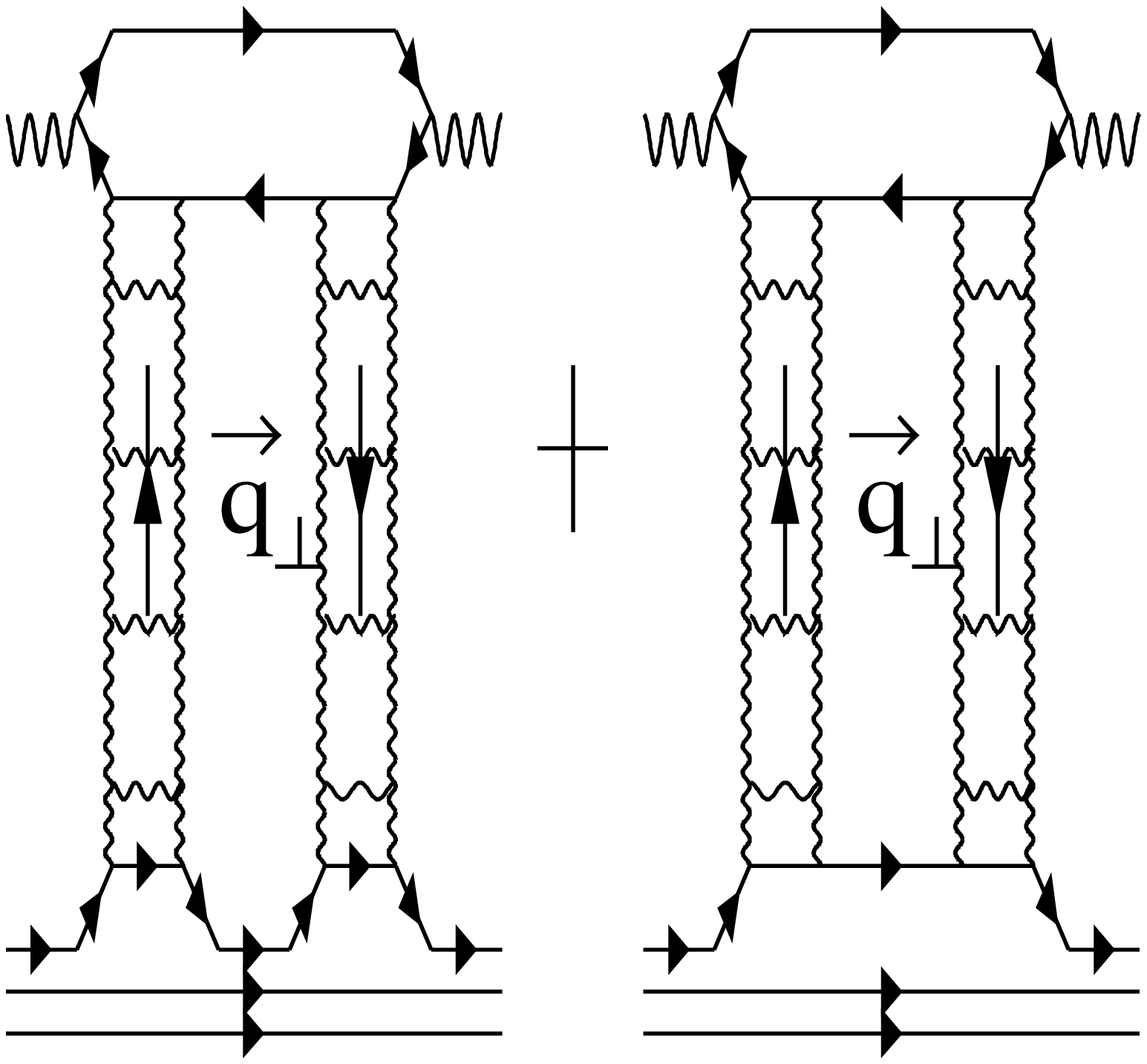,width=140mm,height= 65mm}
\caption{The SC for the total cross section of $\gamma^* p $ interaction.}
\label{Fig.4}     
\end{figure}
Let us discuss this point a little bit in more details. The CDF
collaboration measured the processes of unclusive production of two
pairs of ``hard" jets with almost compensated transverse momenta in each
pair and with almost the same values of rapidities. Such pairs can be
produced only due to double parton collision and their cross section can
be calculated using the Mueller diagram given in Fig.5.

\begin{figure}
\epsfig{file=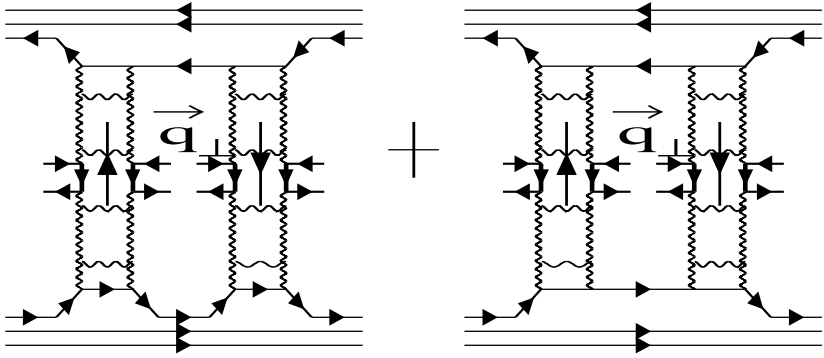,width=140mm,height=65mm}
\caption{Inclusive production of two pair of ``hard" jets in the  double
parton scattering.}
\label{Fig.5}
\end{figure}
 The value of the double parton scattering cross section can be written in
the form ( see Fig.5)\cite{CDF}:
\beq \label{DP}
\s_{DP}\,\,=\,\,m\,\,\frac{\s(\bar Q_1 Q_1) \,\,\s(\bar Q_2
Q_2)}{2\,\s_{eff}}\,\,,
\eeq
where, for simplicity,  we consider the production of two ( $\bar Q_1 Q_1$
and $\bar Q_2 Q_2
$ ) quark - antiquark pairs. Factor $m$ in \eq{DP} is equal to 2 for
different quarks ( $ Q_1\,\neq \,Q_2$) and to 1 for identical quarks. The
 value for $\s_{eff}$ is measured to be 14.5 $\pm$ 1.7 $\pm$ 2.3 mb.
Our estimates \cite{GLMDP} for diagrams of Fig.5 using the two radii
picture give $\s_{eff} \,\sim\,17 mb$. It means that the effective radius 
$R^2 = 5 GeV^{-2}$ could be even overestimated.

 Using the GRV parameterization  for the gluon
structure function and the value of $R^2 = 5 \, GeV^{-2}$,
we obtain  that $\kappa$ reaches 1 at HERA
kinematic region ( see Fig.1 ), meaning shadowing corrections
should not  be neglected. In Fig.1 we plotted three curves with values of
$\kappa$ equal to 1.6, 1 and 0.6, respectively, to illustrate a possible
range of $\kappa$ using CTEQ and MRS parameterizations.

\centerline{}
{\bf A1:}
\centerline{}
The answer to the first question one can read from Fig.1. Indeed, the 
kinematic region where the BFKL Pomeron ( the BFKL corrections to the
anomalous dimension )  could be sizeable, namely, the region between
curves $<\gamma>$ = 1/2 and $\gamma$ = 1 , is located to the left of the
curve with  $\kappa$ = 1 where the SC should be essential. Therefore, we
can conclude that the BFKL Pomeron is hidden under large SC and cannot be
observed. To illustrate this point and to show what is the influence of
the SC on the behaviour of the average anomalous dimension we plotted in
Fig.6  $< \gamma > $ given by Eq.(6) but using the Glauber - Mueller
formula for the SC for the gluon structure function ( see Ref.
\cite{AGLP} ). The calculations were performed for the gluon structure
function at fixed impact parameter ($b_t = 0 $), where we take
$$
x g(x,Q^2) \,=\,\int d^2 b_t S(b_t) x G(x, Q^2)\,\,,$$ with
$$S(b_t)\,\,=\,\,\frac{1}{\pi R^2} e^{-\,\frac{b^2_t}{R^2}}\,\,.$$
One can see, that the average anomalous dimension turns out to be smaller
that $<\gamma>\,=\,\frac{1}{2}$. Therefore, the BFKL Pomeron will not be
seen even if we will take the SC at the minimal rate given the Glauber -
Mueller formula.

\centerline{}
{\bf A2:}
\centerline{}
It turns out that the SC is not very big for  $F_2(Q^2,x)$ which has
been measured experimentally ( see Ref.\cite{AGLP} for details).
However, the SC for $xG(Q^2,x)$ should be large. To illustrate this point
we plot in Fig.6 the ratio $R_1
\,=\,\frac{xG(Q^2,x)^{SC}}{xG(Q^2,x)^{GRV}}$ calculated in Ref.
\cite{AGLP}. Comparing Fig.6 with the value of $xG(Q^2,x)$ in  current
parameterizations we can conclude that in spite of sufficiently large SC 
our knowledge of the value of the gluon structure
function is so poor that we can absorb all SC in the uncertainties 
of its value. For example, taking into account the SC the GRV gluon
structure function will be able to describe the new experimental data
while without the SC the GRV parameterization has been ruled out by
experiment ( see the last picture in Fig.2.).

\begin{figure}[htbp]
\begin{tabular}{l r}
\epsfig{file= 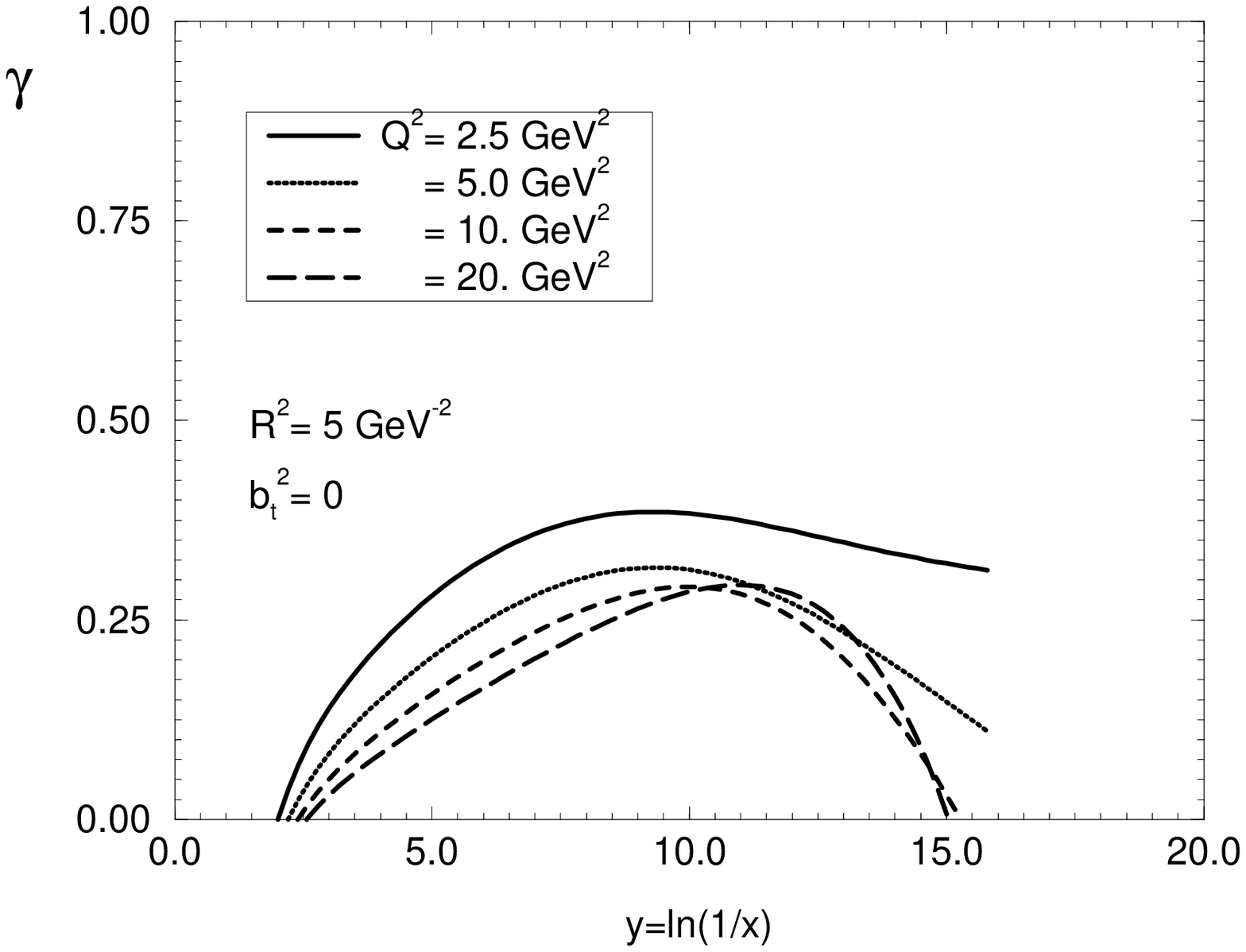,width=70mm} & \epsfig{file= 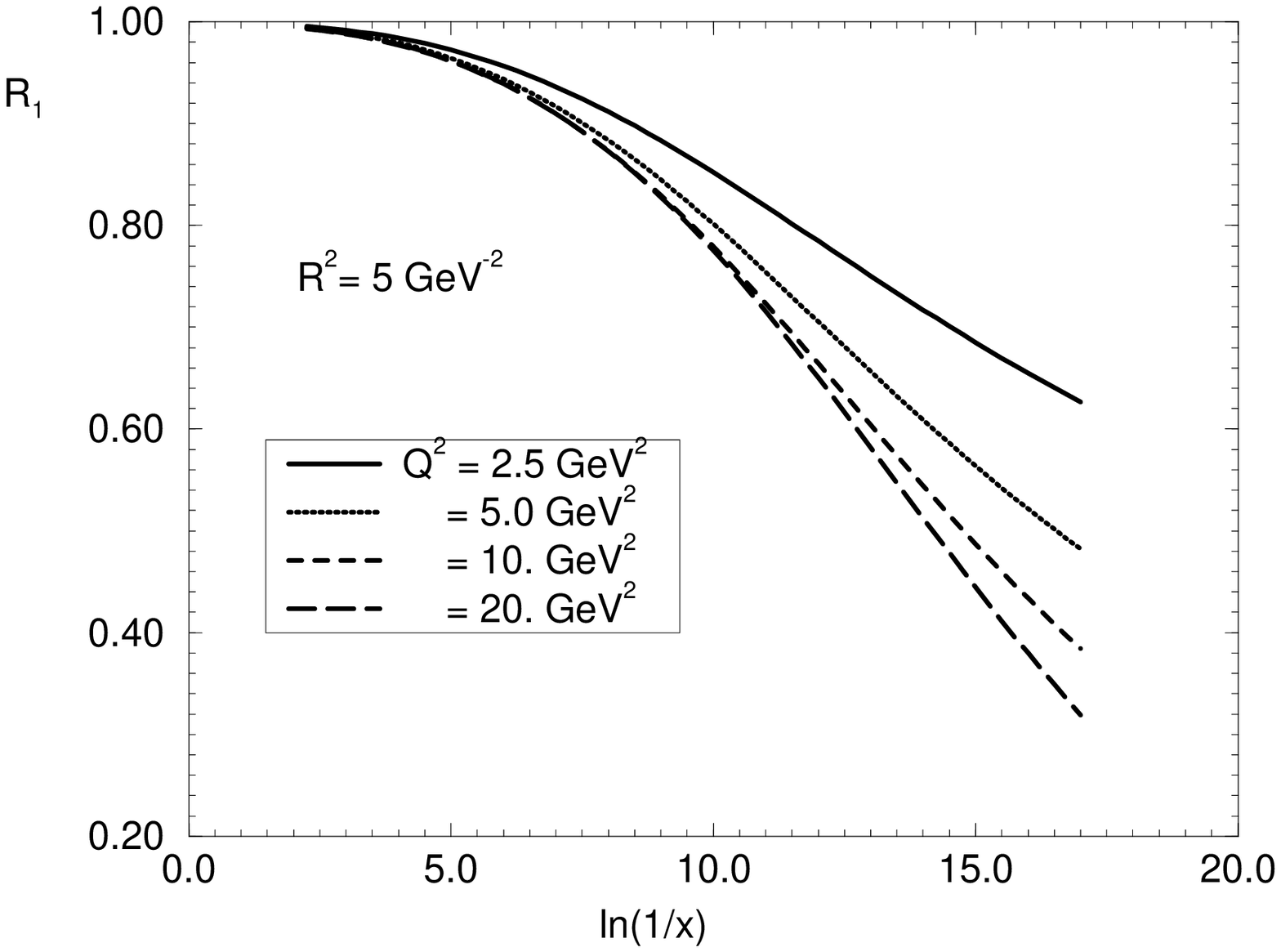,width=70mm}\\ 
\end{tabular}
\caption{ Average anomalous dimension, given in Eq.(6) for the SC taken in
Mueller-Glauber formula for $b_t = 0 $ and
 ratio $R_1$ for the
gluon structure function.} 
\label{Fig.6}
\end{figure}

It should be stressed, that our answers bring several problems that have
to be solved in the nearest future. These are three of them:

{\bf P1:}  

It is very likely that the BFKL Pomeron  will be hidden under SC not only
for the deep inelastic proton structure function but in all specially
invented
processes to extract the BFKL Pomeron since the typical size ( the value
of $R^2$) is smaller in all such processes than in the case of the deep
inelastic structure function. 

{\bf P2:}

The fact that the SC will take place of the BFKL Pomeron does not mean 
that the the experimental cross section will be the same as in the DGLAP
evolution equations or in the Monte Carlo simulations based on the DGLAP
evolution. The difference should be calculated to be discussed.

{\bf P3:}

The size of the SC for the gluon structure function crucially depends on
the initial gluon distribution. In our estimates we pretended that the GRV
parameterization guessed correctly this initial distribution. The only
argument is the fact that the GRV parameterization describes  the
experimental $F_2 (x,Q^2)$ at small $Q^2 \,\approx\, 1 \,GeV^2$.

{\bf Alternative answers:}

{\bf AA1:}

The first alternative answer has been proposed by R. Thorne (see Ref.
\cite{THORNE} ) who demonstrated  that the
correct 
inclusion of the BFKL anomalous dimension allows to improve the comparison
with the experimental data. However, Fig.7 shows that the value of the
gluon structure function extracted from the experimental data is not
very different from the previous analysis without the BFKL anomalous
dimension. Therefore, the experimental data, perhaps, does not contradict
the existence of the BFKL Pomeron, but this fact cannot change
considerably the value of $\kappa$. It means, that the value of the SC is
still big even in the analysis taking into account the BFKL Pomeron ( see
KMS paper \cite{THORNE} for details).  In Fig.7 is given  some
next to leading order corrections to the BFKL Pomeron. One can see that
they are essential and diminish the value of the gluon structure function
but still not more than in two times, which we evaluate as a typical error
in the value of extracted gluon structure function.

\begin{figure}[htbp]
\centerline{\epsfig{file=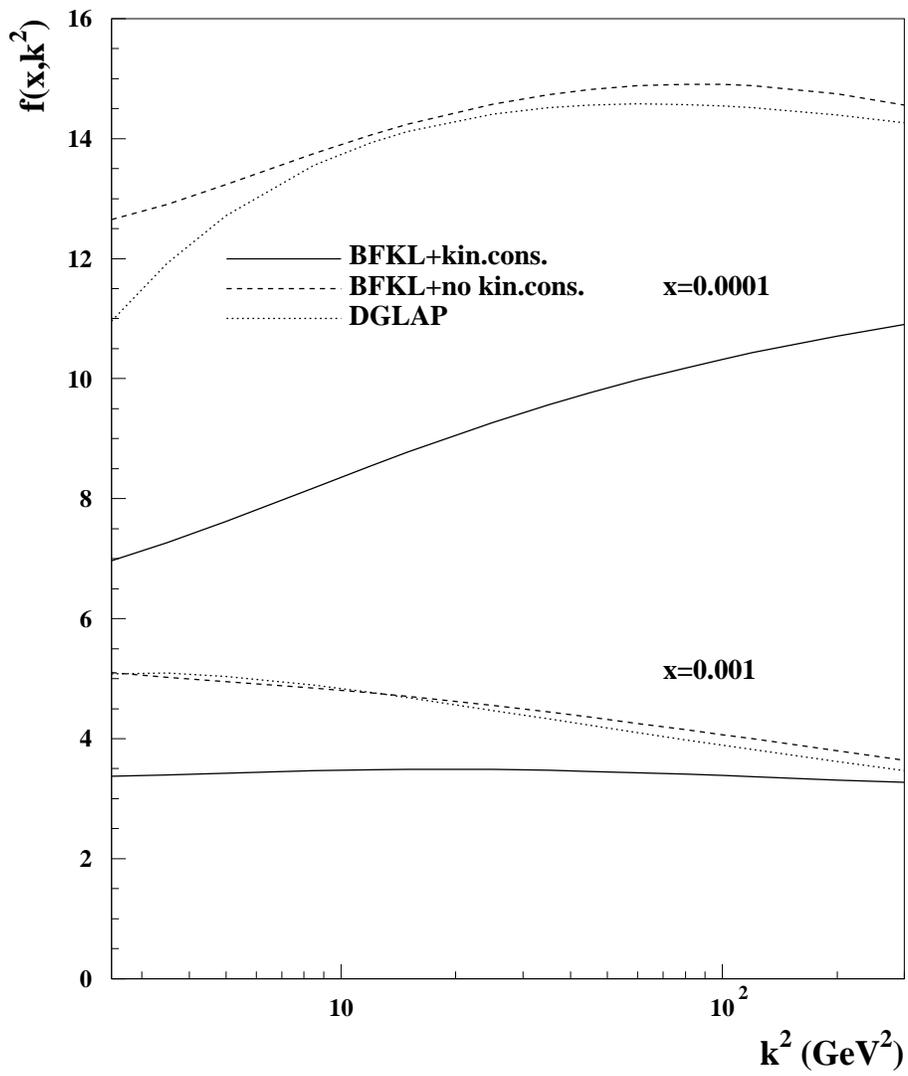,width=140mm}}
\caption{ The gluon structure function with and without the BFKL Pomeron 
 ( picture is taken from the KMS paper \protect\cite{THORNE}).}
\label{Fig.7}
\end{figure}

{\bf AA2:}

The BFKL Pomeron is not seen in the data because the next to leading
correction is essential and they change crucially the main properties of
the BFKL Pomeron. Fortunately, the next order correction to the BFKL
Pomeron (NOBFKL) has been calculated \cite{BFKLNL} and the community of
experts has
started to understand the influence of the NOBFKL \cite{CC1} \cite{BLUM}
on the value of the gluon structure function. It turns out the the NOBFKL
Pomeron has a much smaller intercept or in other word the power - like
behaviour, namely, $xG*x,Q^2)\,\,\propto\,\,(\frac{1}{x_B} )^{\omega_L}$,
 still remains but $\omega_l^{NO}\,\,=\,\,\omega^{LO}_L (\, 1\,\, -\,\,3.4
\,\as\,)  $. It means that the energy ( $x$) dependence becomes milder
and it makes the NOBFKL Pomeron not so pronounced as it was in the leading
order. Nevertheless, the first numerical estimates show that the value of
the gluon structure function changes but not significally. Indeed, Fig.8
that was taken from Ref.\cite{BLUM} shows that the gluon structure
function in the analysis with the NOBFKL anomalous dimension typically on
30\% less than the gluon structure function without the BFKL contribution.

\begin{figure}[htbp]
\centerline{\epsfig{file=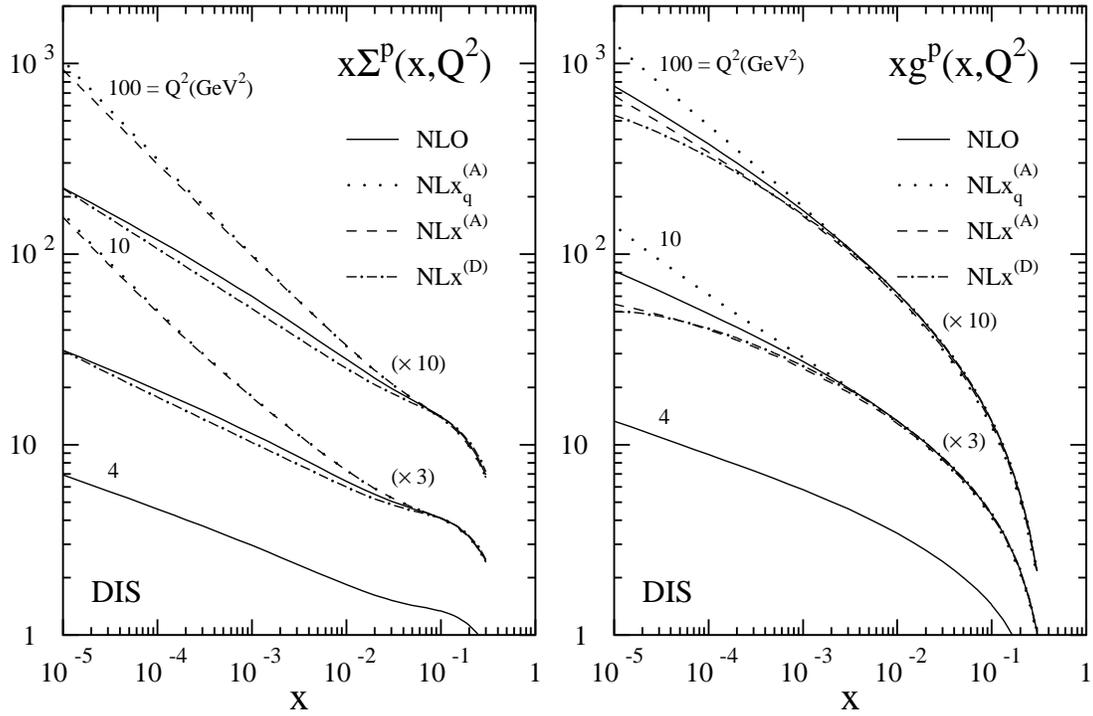,width=150mm}}
\caption{ The gluon structure function with and without the NOBFKL Pomeron
 ( picture is taken from the Blumlein and Vogt  paper
\protect\cite{BLUM}).}
\label{Fig.8}
\end{figure}

{\bf My conclusions:} 

These two examples of the alternative answer show that (i) it is not so
easy to change significally the value of the gluon structure function
 and diminish it more than in two times; and (ii) the real accuracy of the
value of the gluon structure function is rather big, about 50\%, in spite
of the fact that  the difference between two sets of gluon structure
functions ( so called global fits: MRS and CTEQ )  became  much smaller 
using new more accurate experimental data (see Fig.2).  Our errors are
mostly theoretical ones. In my opinion, we cannot change ( diminish ) the
value of parameter $\kappa$ and therefore, accordingly to Fig.1, we have to
deal first with the SC and only after that to take into account the BFKL
Pomeron with all possible corrections.

Finally, I would like to emphasize that you got my personal answers.
Perhaps, you have different ones. The only point, which I insist on, is
that,
before answering these two questions, we cannot trust the DGLAP evolution
more than any other model.

\end{document}

%% file: defi.tex
 1
\def\as{\alpha_{\rm S}}

\def\citenum#1{{\def\@cite##1##2{##1}\cite{#1}}}
\def\citea#1{\@cite{#1}{}}

\def\as{\alpha_{\rm S}}

\def\s{\sigma}

\def\({\left(}
\def\){\right)}

\def\citenum#1{{\def\@cite##1##2{##1}\cite{#1}}}
\def\citea#1{\@cite{#1}{}}

\def\l1vt{\vec{l_{1\perp}}}

\def\rt{r_{\perp}}
\def\bt{b_{\perp}}
\def\rt2{$r^2_{\perp}$}
\def\bt2{$b^2_t$}

\def\jol1{$J_0(\,l_{1\perp}\,r_{\perp}\,)$}

\def\citea#1{\@cite{#1}{}}

\def\beq{\begin{equation}}
\def\eeq{\end{equation}}
\def\bea{\begin{eqnarray}}
\def\eea{\end{eqnarray}}

\def\eq#1{{Eq.~(\ref{#1})}}

\def\bbbz{{\mathchoice {\hbox{$\sf\textstyle Z\kern-0.4em Z$}}
{\hbox{$\sf\textstyle Z\kern-0.4em Z$}}
{\hbox{$\sf\scriptstyle Z\kern-0.3em Z$}}
{\hbox{$\sf\scriptscriptstyle Z\kern-0.2em Z$}}}}
\def\l{\lambda}